\documentclass[manuscript]{aastex}

\slugcomment{Not to appear in Nonlearned J., 45.}

\shorttitle{Inner disk of J16042165-2130284}
\shortauthors{Mayama et al.}

\begin{document}

\title{ALMA Reveals a Misaligned Inner Gas Disk\\inside the Large Cavity of a Transitional Disk
    }

\author{Satoshi Mayama\altaffilmark{1}, 
Eiji Akiyama\altaffilmark{2},
Olja Pani\'c \altaffilmark{3}$\dagger$,
James Miley\altaffilmark{3},
Takashi Tsukagoshi\altaffilmark{4},
Takayuki Muto\altaffilmark{5},
Ruobing Dong\altaffilmark{6},
Jerome de Leon\altaffilmark{7},
Toshiyuki Mizuki\altaffilmark{8},
Daehyeon Oh\altaffilmark{9},
Jun Hashimoto\altaffilmark{10},
Jinshi Sai\altaffilmark{7},
Thayne Currie\altaffilmark{11},
Michihiro Takami\altaffilmark{12},
Carol A. Grady\altaffilmark{13,14,15},
Masahiko Hayashi\altaffilmark{4},
Motohide Tamura\altaffilmark{4,7,10},
Shu-ichiro Inutsuka\altaffilmark{16}
}

\altaffiltext{1} {SOKENDAI(The Graduate University for Advanced Studies), Shonan International Village, Hayama-cho, Miura-gun, Kanagawa 240-0193, Japan}
\email{mayama\_satoshi@soken.ac.jp}
\altaffiltext{2} {Institute for the Advancement of Higher Education, Hokkaido University, Kita 8, Nishi 5, Kita-ku, Sapporo, Hokkaido 060-0808, Japan}
\altaffiltext{3} {School of Physics and Astronomy, University of Leeds, Woodhouse Lane, Leeds LS2 9JT, UK}
\altaffiltext {4} {National Astronomical Observatory of Japan, National Institutes of Natural Sciences, 2-21-1 Osawa, Mitaka, Tokyo 181-8588, Japan}
\altaffiltext{5} {Division of Liberal Arts, Kogakuin
University, 1-24-2, Nishi-Shinjuku, Shinjuku-ku, Tokyo, 163-8677, Japan}
\altaffiltext{6} {Department of Physics \& Astronomy, University of Victoria, 3800 Finnerty Rd, Victoria, BC V8P 5C2, Canada}
\altaffiltext{7} {Department of Astronomy, The University of Tokyo,
Hongo 7-3-1, Bunkyo-ku, Tokyo 113-0033, Japan}
\altaffiltext{8} {Institute of Space and Astronautical Science, JAXA, 3-1-1 Yoshinodai, Chuo-ku, Sagamihara, Kanagawa 252-5210, Japan}
\altaffiltext{9} {National Meteorological Satellite Center, 64-18 Guam-gil, Gwanghyewon-myeon, Jincheon-gun, Chungbuk, South Korea}
\altaffiltext {10} {Astrobiology Center, National Institutes of Natural Sciences, 2-21-1 Osawa, Mitaka, Tokyo 181-8588, Japan}
\altaffiltext{11} {Subaru Telescope, 650 North A'ohoku Place, Hilo, HI
96720, USA}
\altaffiltext {12} {Institute of Astronomy and Astrophysics, Academia
Sinica, P.O. Box 23-141, Taipei 106, Taiwan}
\altaffiltext {13} {Goddard Center for Astrobiology, NASA's Goddard Space Flight Center, Greenbelt, MD 20771, USA}
\altaffiltext {14} {Eureka Scientific, 2452 Delmer, Suite 100, Oakland CA
96002, USA}
\altaffiltext {15} {ExoPlanets and Stellar Astrophysics Laboratory, Code
667, Goddard Space Flight Center, Greenbelt, MD 20771 USA}
\altaffiltext{16} {Department of Physics, Nagoya University, Furo-cho, Chikusa Ward, Nagoya, Aichi 464-0814, Japan}
\altaffiltext{$\dagger$} {Royal Society Dorothy Hodgkin Fellow}




\begin{abstract}

Pairs of azimuthal intensity decrements at near symmetric locations have been seen in a number of protoplanetary disks. They are most commonly interpreted as the two shadows cast by a highly misaligned inner disk. Direct evidence of such an inner disk, however, remain largely illusive, except in rare cases. In 2012, a pair of such shadows were discovered in scattered light observations of the near face-on disk around 2MASS J16042165-2130284, a transitional object with a cavity $\sim$60 AU in radius. The star itself is a ``dipper'', with quasi-periodic dimming events on its light curve, commonly hypothesized as caused by extinctions by transiting dusty structures in the inner disk. Here, we report the detection of a gas disk inside the cavity using ALMA observations with $\sim0$\farcs2 angular resolution. A twisted butterfly pattern is found in the moment 1 map of CO (3-2) emission line towards the center, which is the key signature of a high misalignment between the inner and outer disks. In addition, the counterparts of the shadows are seen in both dust continuum emission and gas emission maps, consistent with these regions being cooler than their surroundings. Our findings strongly support the hypothesized misaligned-inner-disk origin of the shadows in the  J1604-2130 disk. Finally, the inclination of inner disk would be close to -45 $^{\circ}$ in contrast with 45 $^{\circ}$; it is possible that its internal asymmetric structures cause the variations on the light curve of the host star.

\end{abstract}



\section{Introduction}

Planets are believed to form in protoplanetary disks. As they do, they gravitationally perturb their host disk, and generate structures \citep[and reference therein]{kley12}. In recent years, a variety of features have been discovered in spatially resolved observations of protoplanetary disks, such as spiral arms \citep[e.g.,][]{fukagawa06, hashimoto11, muto12, 2015ApJ...814L..27C}, gaps or cavities \citep[e.g.,][]{hashimoto12, brogan15, 2015ApJ...802L..17A, tsukagoshi16}, and dust traps \citep[e.g.,][]{vandermarel13, casassus13}. Such structures may be produced by massive planetary companions forming in disks \citep[e.g.,][]{dong15gap, dong15spiralarm}. By studying planet-induced structures and comparing simulations with observations, it is possible to infer properties of the feature-producing planets, such as their masses \citep[e.g.,][]{dong17gap, dong17spiralarm}.

One specific disk, which is the focus of this paper, is shadows. They have been seen in disk images at both near-infrared (NIR) and millimeter (mm) wavelengths, and are generally thought to be the outcome of starlight being obscured by structures in the inner disk. A prototype of such features was discovered in the HD 142527 disk, which shows two narrow nulls at near-symmetric locations ($m\sim2$) on the disk ring at $\sim$100 AU in scattered light \citep{fukagawa06, 2012ApJ...754L..31C, canovas13, avenhaus14, rodigas14}. The counterparts of the two nulls in dust continuum emission were subsequently discovered in ALMA observations as two local depressions in mm surface brightness \citep[see also \citealt{muto15}]{2015ApJ...812..126C}. \citet{2015ApJ...798L..44M} proposed that the two nulls in scattered light were shadows casted by an inner disk $\sim$10 AU in radius and highly misaligned from the outer disk ($\sim70^\circ$ mutual inclination). Their model naturally explains the mm observations as well -- the regions in the shadows are cooler than their surroundings as the stellar heating is blocked, resulting in a reduction in dust emission \citep{2015ApJ...812..126C}. 
A review about warps in transition disks can be found in \citet{2016PASA...33...13C}.

Here, we introduce our target 2MASS J16042165-2130284 (hereafter J1604-2130) in Upper Scorpius. Recent GAIA DR2 reported that J1604-2130 is located at a distance of 150 pc. 
\citet{2012ApJ...746..154P} derived stellar age 10~Myr, an epoch that probes the final stages of jovian planet formation.
\citet{1999AJ....117.2381P} derived spectral type K2, stellar mass 1~$M$$_\odot$, $\log T_\mathrm{eff}$[K]=3.658, and $\log[L/L~_\odot]=-0.118$.  
J1604-2130 has mass accretion rate of no larger than 10$^{-11}$~M$_\odot$ yr$^{-1}$ \citep{2012ApJ...753...59M,2009AJ....137.4024D}.  
\citet{2012ApJ...760L..26M} detected two shadows on the disk in Subaru near-infrared scattered light imaging observations. Although they reported the detection of an arc-like structure, the team later determined it to be an artifact.

Lastly, J1604-2130 is known to have two different issues of variability. Firstly, J1604-2130 has been identified as a ``dipper'' \citep{2016MNRAS.462L.101A, ansdell16dipper} --- young stellar objects that exhibit quasi-periodic or aperiodic dimming on their optical and infrared light curves. Secondly, while the Spitzer IRAC NIR photometry \citep{carpenter06} shows an SED well consistent with the stellar photosphere, the Spitzer IRS mid-infrared (MIR) spectrum \citep{2009AJ....137.4024D} and WISE photometry, taken at different times, show NIR-to-MIR excess. The latter data suggest the presence of a sub-AU inner disk (see the discussion in \citealt{zhang14}). These two issues may be related, but not the same issue.

In this paper, we report a new high-resolution ALMA observation of both dust and gas in the J1604-2130 system. Our goal is to test the hypothesis that the two shadows seen in scattered light imaging are cast by an inner disk highly misaligned with the outer disk, by searching for depressions in mm surface brightness at the location of the shadows and the twisted butterfly pattern in gas moment 1 maps, both being key signatures of such a disk structure.

\section{Observations and Data Reduction}

J1604-2130 was observed in ALMA cycle 3 program 2015.1.000888.S (PI: E. Akiyama) in Band 7. The observations were conducted with the extended and compact configurations. The observation with the compact configuration was performed on March 10, 2016 with 37 operative antennas, and the observations with the extended configuration were performed on August 2 and 15-16, 2016 with 40 operative antennas.  The extended configuration array provided a maximum and minimum baseline length of 1.1 km and 15.1 m, respectively. The compact configuration array provided a maximum and minimum baseline length of 460.0 m and 15.3 m, respectively. Sky conditions were relatively stable for 1.3 mm wavelength observations with the precipitable water vapor between 0.50 mm and 1.06 mm. During observation, the system temperatures were between 100 K and 300 K. The ALMA correlator was set to have two continuum windows and two spectral windows. One spectral window with bandwidth of 234.375 MHz was centered at 345.8 GHz for the CO (3-2) (channel widths of 244.141 kHz; equivalent to a velocity resolution of $\sim$211.68 m s$^{-1}$).  Another spectral window with bandwidth of 468.750 MHz was centered at 356.7 GHz for the HCO$^{+}$ (4-3) (channel widths of 488.281 kHz; equivalent to a velocity resolution of $\sim$410.37 m s$^{-1}$). The other windows with bandwidth of 2.00 GHz were configured to obtain the continuum emission centered at 344.0 GHz and 355.5 GHz (channel widths of 15.625 MHz). Bandpass was calibrated using J1517-2422. Absolute flux density was scaled using J1517-2422 or Titan.  J1625-2527 or J1553-2422 was observed for phase calibration. The total integration time on source for target J1604-2130 was 62.8 minutes with extended configuration and 31.4 minutes with compact configuration.

The data were calibrated using version 4.7 of the Common Astronomy Software Applications package (CASA) for extended configuration data and version 4.5 for compact configuration data. We followed the pipeline calibration provided by East Asian ALMA Regional Center and as provided for this data set on the ALMA archive. Because the line and continuum observations were simultaneously conducted with adjacent frequencies, and the atmospheric variation equally affects the phase variation of line and continuum data, the calibration table obtained from the continuum data was applied to line emission calibration. As we concatenate all the observations together for maximum uv-coverage and present/analyze the resulting continuum map, we show only the combined continuum map (low and high resolution) in further sections. The same applies for HCO$^{+}$ (4-3) and CO (3-2). We used the clean algorithm in CASA for imaging. The self-calibration was performed by applying the obtained continuum image as a model, and gain calibration was repeated until the rms converged to a minimum. The gain table obtained after the self-calibration of the continuum data was applied to both the CO (3-2) and HCO$^{+}$ (4-3) lines data to generate self-calibrated visibilities. The Briggs weighting with robust of 0.5 was applied in both continuum and line imaging to obtain the optimal combination of resolution and image fidelity. 
The achieved rms in the continuum image using in all spectral windows excluding the line emission channels was 39 $\sim \mu$Jy beam$^{-1}$. The achieved rms in the individual channel maps with approximately native channel width was 3.813 mJy beam$^{-1}$ for CO (3-2), and 3.385 mJy beam$^{-1}$ for HCO$^{+}$ (4-3). The achieved rms of moment 0 map was 6.454 mJy/beam.km/s for CO (3-2), and 6.965 mJy/beam.km/s for HCO$^{+}$ (4-3). The final achieved synthesized beam of approximately 0.$''$216 $\times$ 0.$''$187 (PA=76.7 $^{\circ}$), 0.$''$229 $\times$ 0.$''$194 (PA=77.8 $^{\circ}$) and 0.$''$218 $\times$ 0.$''$188 (PA=72.1 $^{\circ}$) after calibration and flagging for the continuum, CO (3-2) and HCO$^{+}$ (4-3), respectively. As the following section shows, a disk is detected in HCO$^{+}$ (4-3), CO (3-2) emission and continuum. For the stellar position, we used the GAIA DR2.


\section{Results}

\subsection{HCO$^{+}$ (4-3) Emission} \label{bozomath}

\begin{figure}
\includegraphics[angle=0,scale=1.0]{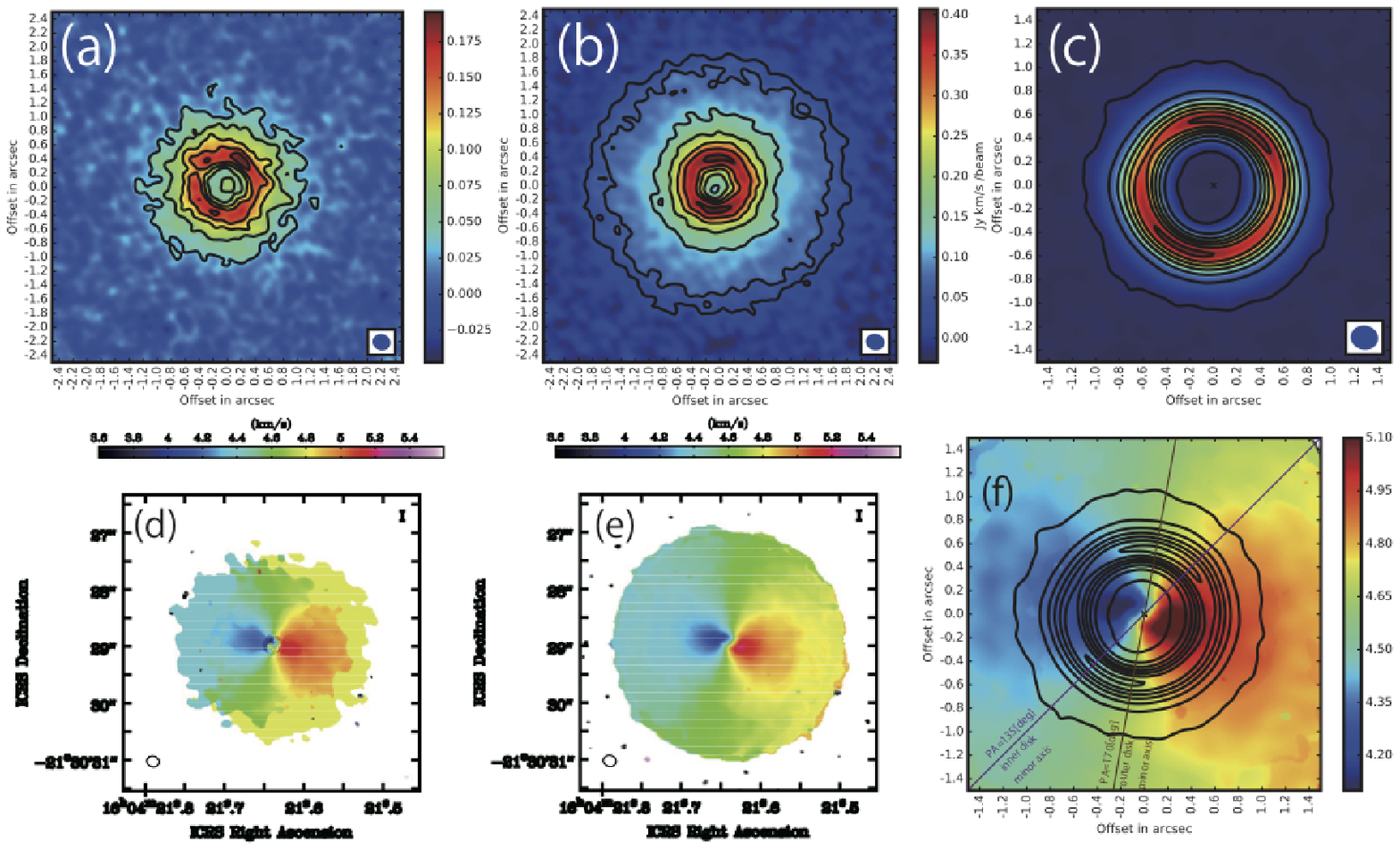}
\caption{ALMA images of J1604-2130. An ellipse at the bottom right corner for (a),(b),(c), and bottom left corner for (d),(e) denotes the ALMA synthesized beam. The unit of the color bar for (a), (b) and (d),(e),(f) is [Jy/beam.km/s] and [km/s], respectively.   (a) HCO$^{+}$ (4-3) moment 0 map.  Contour levels are (5,10,15,20,25)$\times$rms. (b) CO (3-2) moment 0 map. Contour levels are (5,10,20,30,40,50,60)$\times$ rms. (c) Color map of continuum emission overlaid with and contours at (5,50,100,150,200,250,300)$\times$rms. (d)HCO$^{+}$ (4-3) moment 1 map. (e)CO (3-2) moment 1 map. (f)CO moment 1 map is shown in the color map.   Continuum in black contours at (5,50,100,150,200,250,300)$\times$rms is overlaid. Purple color line denotes the position angle 135$^{\circ}$ of inner disk minor axis. Brown color line denotes the position angle 170$^{\circ}$ of outer disk minor axis.The black cross gives the stellar position. 
}
\label{fig1}
\end{figure}

Figure \ref{fig1} shows the moment 0 map, integrated intensity, for HCO$^{+}$ (4-3) emission around J1604-2130. Figure \ref{fig4} (a) shows an azimuthally averaged normalized intensity. As clearly seen in this figure \ref{fig4}(a), the HCO$^{+}$ (4-3) emission radial profile has its first bright peak at r=0$''$ from the central star, then becomes fainter around the cavity zone, then have its second bright peak at the outer ring zone, and then becomes fainter again.  In other words, emission extends into the cavity with a dip in the profile between a peak at the center of the cavity and bright outer ring. The moment 0 map in figure \ref{fig1} also shows a peak at the center of the cavity. This inner component has 12.4 $\sigma$ and its size is close to the beam size. This reveals that there are still materials inside this cavity.  
Another salient feature detected in the moment 0 map are two local depressions, or dips, in the emission in the east-northeast and west-southwest side of the disk. These dips are also seen in figure \ref{fig4} (b) and their origins are discussed in section $\S$\ref{sect:dips}. 

Figure \ref{fig1}(d) shows the moment 1 map, which shows mean velocity for HCO$^{+}$ (4-3) emission in the J1604-2130 disk. The central velocity, 4.6 km/s for J1604-2130, traces the material along the minor axis in disks as can be seen in green color in figure \ref{fig1}(d).

Figure \ref{fig5} displays the HCO$^{+}$ (4-3) spectrum obtained by integrating over a circular area centered at the location of J1604-2130. The central velocity, 4.6 km/s is seen as a peak in this figure. The HCO$^{+}$ (4-3) integrated spectra over the inner component shows a double peak. Based on these figures, the inner component peak at the center of the cavity is interpreted as HCO$^{+}$ (4-3) inner disk. 

\subsection{CO (3-2) Emission} \label{bozomath}

Figure \ref{fig1} shows the moment 0 map, integrated intensity, for CO (3-2) emission of the J1604-2130. In this map, the faintest part in the ring has an offset and is located slightly south of the central star. The CO (3-2) emission particularly in the inner part shows a different view from that of HCO$^{+}$ (4-3) probably due to a difference in their optical depth. There are also two dips detected in the CO (3-2) moment 0 map.  Orientation of the east dip is slightly different from HCO$^{+}$ (4-3) or continuum.
These dips are also seen in figure \ref{fig4} (b) and their origins are discussed in section $\S$\ref{sect:dips}.  

Figure \ref{fig1}(e) shows the moment 1 map, which is intensity-weighted velocity, for CO (3-2) emission of the J1604-2130. The CO (3-2) moment 1 map shows a twisted kinematic structure in the central area.  These twisted first-moment maps have already been observed in several sources by ALMA
\citep[e.g.,][]{2015ApJ...811...92C,2017A&A...607A.114W,2017A&A...597A..32V,2014ApJ...782...62R}. In the moment 1 map for CO (3-2) emission image, two rotational components can be interpreted as CO (3-2) inner disk and outer disk inferred from twisting structures.
CO (3-2) spectrum obtained by integrating over a circular area centered at the location of J1604-2130 is displayed in Figure \ref{fig5}. 

We used the CO (3-2) intensity weighted velocity (first moments) map to derive position angles for both outer and inner disks because kinematics provides a finer spatial detail of the morphology (PA) of the rest velocity emission, which corresponds to the minor axis. The spatial resolution on the other hand is not sufficient to measure the inner disk morphology in detail. 
As the zero projected velocity line runs along the semi-minor axis of a disk, based on CO (3-2) moment 1 map shown in figure \ref{fig1} (f), we estimated the P.A. of J1604-2130 outer and inner disk to be 80 and 45$^{\circ}$, respectively. This method to use the central velocity(4.6 km/s) for determination of the PA is described by \citet{2018A&A...615L..10M}.   
The outer disk P.A. 80$^{\circ}$ is consistent with the ALMA cycle 2 result by \citet{2017ApJ...836..201D}. 

\begin{figure}
\includegraphics[angle=0,scale=0.7]{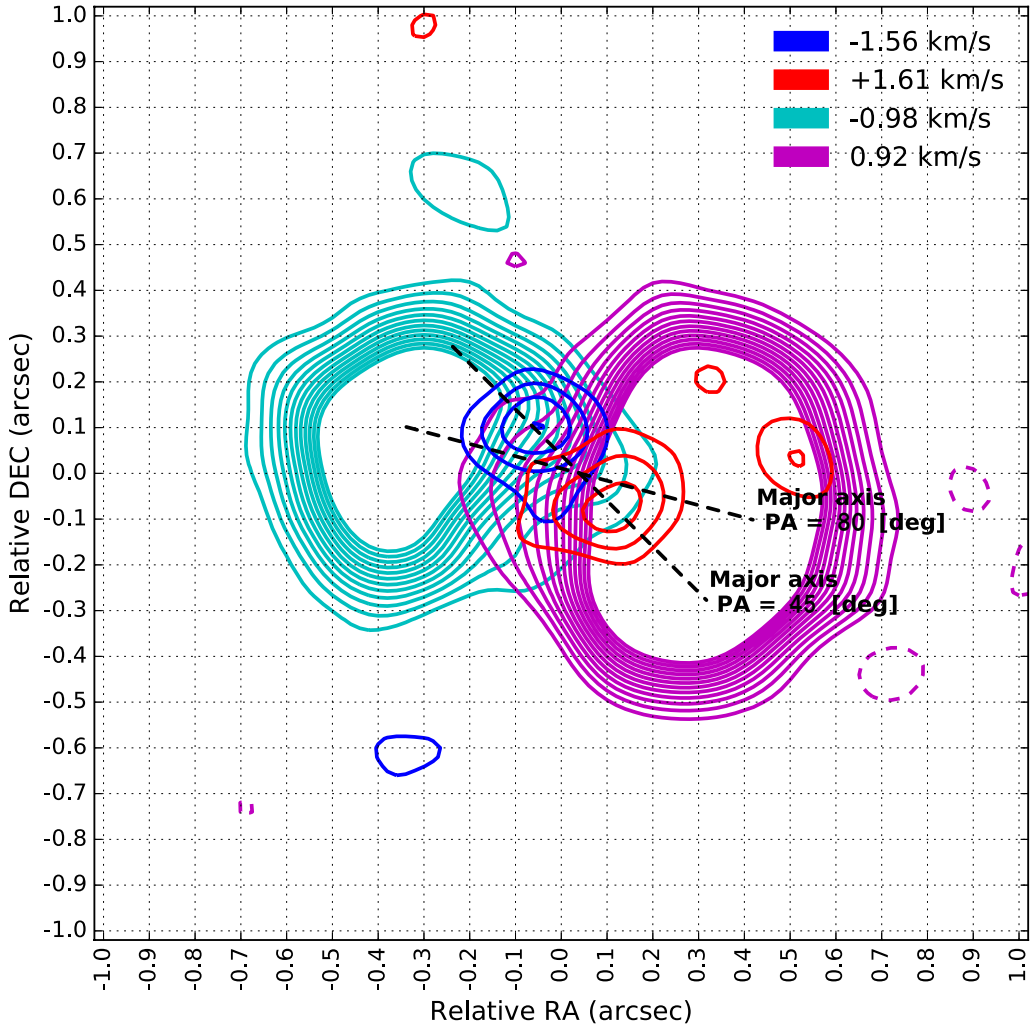}
\caption{A plot of overlaid CO channel maps indicating position angle of the inner and outer parts of the disk. Plotted in blue and red contours are the highest velocity channels containing significant ($>$ 5$\sigma$) emission, probing inner disk regions. Lower velocity channels are plotted in cyan and purple, tracing the outer disk. Dashed lines show the position angle of the inner and outer disk. Selected channels are symmetric around the system velocity within spectral resolution. Contours for each channel begin at 3 sigma, increasing in steps of 2$\sigma$. Cyan and purple contours are plotted up to 25 $\sigma$, blue contours peak at 9 $\sigma$, red peak at 7 $\sigma$, where sigma= 3.8mJy/beam.
}
\label{fig3}
\end{figure}

\begin{figure}
\includegraphics[angle=0,scale=0.9]{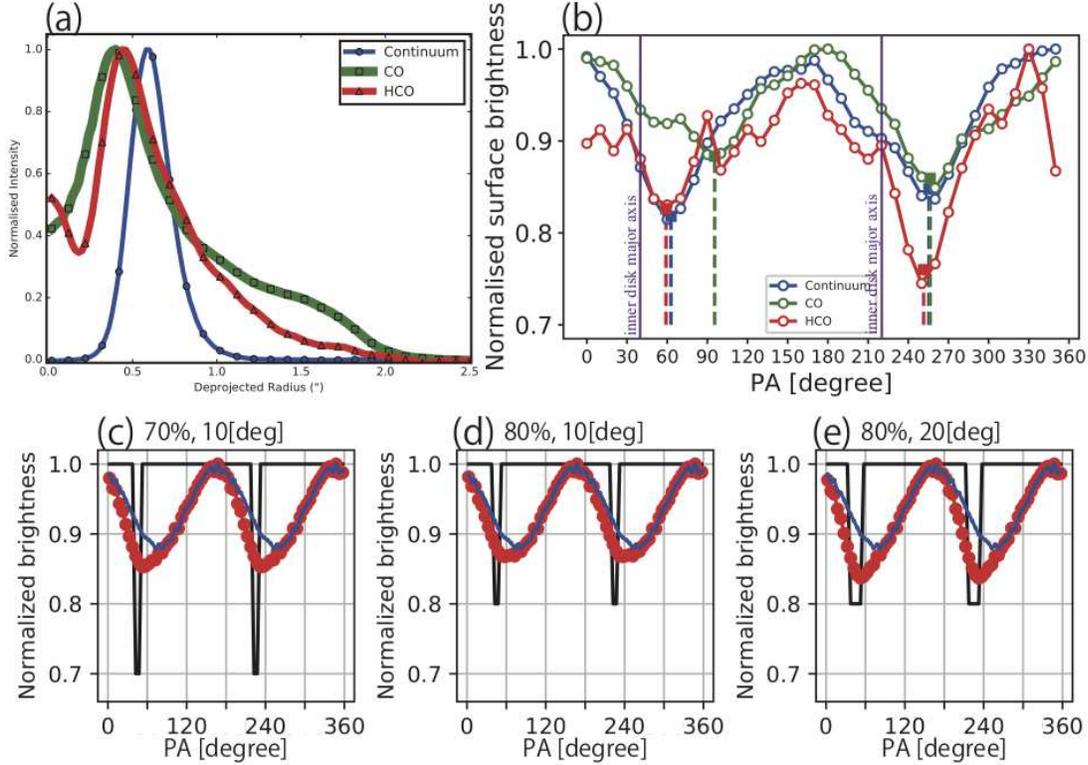}
\caption{(a) Azimuthally averaged normalized intensity of ALMA observations of J1604-2130. Blue, green, and red tracers corresponds to 0.87 mm dust continuum, CO (3-2), and HCO$^{+}$ (4-3), respectively. Normalised intensity against deprojected radius in arcseconds is plotted all assuming PA=80$^{\circ}$, inclination=6$^{\circ}$ \citep{2017ApJ...836..201D}. (b) Azimuthal normalised surface brightness profile of 0.87 mm dust continuum, CO (3-2), and HCO$^{+}$ (4-3) with position angle measured from north to east. Each curve shows the measured flux of pixels. PAs of local minimum fitted by gaussian and PAs of inner disk are annotated. (c),(d),(e) Simulated azimuthal normalized surface brightness profiles by using toy models of a ring disk. Shape of dips including their depth and azimuth width is varied in (c),(d), and (e). 70.0\%, 80.0\%, and 80.0\% labeled on the top of panels denote the dip depth in (c),(d), and (e), respectively. 10 deg, 10 deg, and 20 deg labeled on the top of panels denote the dip azimuth width in (c),(d), and (e), respectively. The black line denotes shape of dips. The red line denotes profile with dips after convolution with an elongated beam same as our ALMA observation. The blue line denotes profile without dips after convolution with an elongated beam same as our ALMA observation.
}
\label{fig4}
\end{figure}

\begin{figure}
\includegraphics[angle=0,scale=1.8]{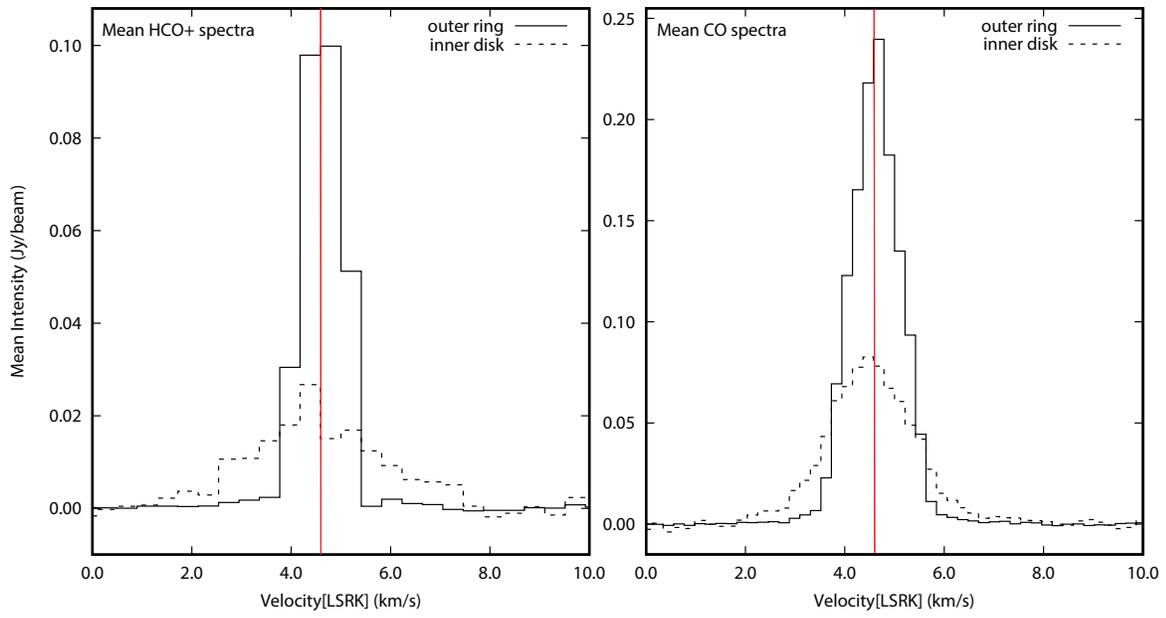}
\caption{(Left) HCO$^{+}$ (4-3) and (Right) CO (3-2) integrated spectra over the entire disk and inner disk (dashed line) of J1604-2130. Central velocity 4.6 km/s is annotated with red line.
}
\label{fig5}
\end{figure}

Figure \ref{fig3} shows a plot of overlaid CO (3-2) channel maps indicating position angle of the inner and outer parts of the disk. With this figure, we confirmed that the PAs of both inner and outer disks derived from CO (3-2) moment 1 map are perfectly consistent with those derived from this CO (3-2) channel map.

\subsection{Continuum Emission} \label{bozomath}

 Figure \ref{fig1} shows the 0.87 mm continuum image of the J1604-2130 with rms of 39 $\mu$Jy beam$^{-1}$.  
As seen in this figure, at lower level, 5 $\sigma$ contours show that the emission is considerably more extended from the star, while the ring is located about 0.$''$6 inwards, the emission extends down to about 0.$''$2 from the star, especially on the east side. This means that it is not empty inside the ring in the continuum as well although the stellar position is free of detectable continuum emission.  Overall, the east side of the ring appears to be broader radially.  There is also an emission signal outside of the ring-like disk.  The Continuum disk also shows two dips. Orientations of these two dips, which are east-northeast and west-southwest, are almost the same as seen in the HCO$^{+}$ (4-3) emission. 

\subsection{Elliptical fitting} \label{bozomath}

  Elliptical fitting was performed to measure outer disk angular separations of the major and minor axes.
  The position angle, derived from CO (3-2) observation, is used as fixed parameter. 
  For the angular separations of the major and minor axes, CO (3-2), HCO$^{+}$ (4-3), and continuum data are separately used for fitting.
We first measured disk radial profiles in 10$^{\circ}$ position angle increments, and then extracted coordinates of the brightest peak area.  
 Those coordinates were used to fit an ellipse and used to measure the surface brightness to plot figure \ref{fig4}(b). The Trust Region Reflective algorithm was performed. The derived semimajor and semiminor axes are called ``peak radii'' in this paper and relatively larger than a cavity wall radius that is often derived in sub mm transition disk modeling studies.  Elliptical fitting results are as follows: for the continuum, peak radii of major and minor axis are $583 \pm 3$ and $576 \pm 3$ [mas], respectively. For HCO$^{+}$ (4-3), peak radii of major and minor axis are $452 \pm 5$ and $418\pm 5$ [mas], respectively. For CO (3-2), peak radii of major and minor axis are $385 \pm 3$ and $375\pm 3$[mas], respectively.


\section{Discussion}

Flux is measured above 3 $\sigma$ from the continuum image giving an integrated emission of 0.276 Jy. Assuming optically thin dust emission and using an opacity for mm-sized particles, observed at a wavelength of 0.87mm = 3cm$^2$/g \citep{2006ApJ...636.1114D}, we assume a minimum disk temperature of 20K, resulting in a derived dust mass lower limit of M$_{dust}$=0.21 M$_\mathrm{Jup}$.  
\citet{2012ApJ...753...59M} derived that J1604-2130 has a dust mass of 0.1~$M$$_\mathrm{Jup}$ by their SMA observation.

\subsection{inner disk geometry and origins of dips} \label{sect:dips}
Figure \ref{fig4}(b) denotes azimuthal normalized surface brightness profile of 0.87 mm dust continuum, CO (3-2), and HCO$^{+}$ (4-3) with position angle measured from north to east. 
By fitting the gaussian profile to the azimuthal profile, the best-fit parameter is found for the location of the local minimum to identify dip PA. The Dip on the east side(east dip) has a PA of 62.0$\pm$0.2, 96.1$\pm$0.5, 60.3$\pm$0.6$^{\circ}$ and the dip on the west side(west dip) has a PA of 255.5$\pm$0.8, 257.0$\pm$0.9, and 251.5$\pm$0.2$^{\circ}$, for dust continuum, CO (3-2), and HCO$^{+}$ (4-3), respectively.  These two dips are the first and second faintest dips for all three tracers. For dust continuum and HCO$^{+}$ (4-3), there is around 190$^{\circ}$ difference between the east dip and west dip. For CO (3-2), there is a difference of around 160$^{\circ}$ between them.  Therefore, these two dips can be called symmetrical dips. 

Very gradual variations of the brightness azimuthally affect the location of the decrements, which are shifted azimuthally because of the finite beam. In order to estimate the impact of this shift and depth of local decrements, we simulate observations using a toy model of a ring disk(=0.$''$58 in radius). 0.$''$1 ring radial width is adopted as a fixed value. Two dips are created at PA of 45 and 225$^{\circ}$, which are theoretically predicted by inner disk PA\citep{2017A&A...605A..16F}. As a result, simulated azimuthal normalized surface brightness profiles with PA measured from north to east are displayed in figure \ref{fig4}(c), (d), and (e). \newline
Regarding a depth of local decrements, this simulation demonstrates the following three points. Firstly, beam elongation at the time of our observation (0.$''$216 $\times$ 0.187$''$ (PA=76.7 $^{\circ}$)) can produce ${\sim}$10${\%}$ of local decrements in surface brightness even without any dips on a disk. Secondly, the depth of these local decrements become deeper by creating dips in a toy model as seen in the figure \ref{fig4}(c), (d), and (e) by comparing red and blue line. Thirdly, we tested that the depth of these local decrements also varies depending on radial width of a ring disk.  \newline
Regarding an azimuth shift, this model demonstrates the following three tendencies. Firstly, locations of observed local decrements can be shifted maximum $\sim$30$^{\circ}$ toward counterclockwise direction from the location predicted by inner disk PA. Secondly, the deeper the dips become, the closer the location of a decrement shifts toward a dip after convolution. Thirdly, the wider the dips becomes azimuthally, the closer the location of a decrement shifts toward a dip after convolution. \newline
In our observation, both the 1st and 2nd faintest continuum decrements have 17 and 30.5 $^{\circ}$ differences from the location predicted by inner disk PA.  These two have different values each other and both are almost within 30 $^{\circ}$ maximum shift toward the same counterclockwise direction derived from this toy model. Therefore, as our interpretations of this toy model, an original existence of continuum dips are required to reproduce these radio decrements and beam elongation can partly contribute both to shift and to enhance dips, which are then consequently seen as observed positions. However, it is not possible to distinguish this interpretation from a case that there is in fact a real deep dip at observed location.

In NIR, similar dips are detected around a P.A. 85 and 255$^{\circ}$ at H-band \citep{2012ApJ...760L..26M}.  Locations of NIR dips are closer to CO (3-2) dips rather than dust continuum and HCO$^{+}$ (4-3). This consistency might be due to NIR and CO (3-2) both being tracing surface layers of the disk. 

\citet{2017A&A...605A..16F} suggests that these dips can be interpreted as two shadows cast by the inner disk onto the outer regions.  This is a direct effect of the dust temperature being lower in the shadowed regions (the column density is close to being azimuthally symmetric in the outer disk).  The stages of the 3D SPH simulations shown in \citet{2017A&A...605A..16F} shared common dips numbers and azimuthal directions with the observed images in the continuum, CO (3-2), and near-infrared.  
These agreements between observations and simulation suggest that two dips are seen by shadowing effects. Moreover, \citet{2017A&A...605A..16F} suggest that the twisted structure in the first-moment map is a peculiar indicator of broken (inner and outer separated) disks rather than warp disk addressed in \citet{2017MNRAS.466.4053J}. 

\citet{2017A&A...604L..10M} probed inner disk geometries from locations of shadows seen in scattered light images by connecting them. Their results as well as results from \citet{2018MNRAS.477.5104C} showed that the position angle of the line connecting the shadows can be related to the position angle of the inner disk. If we adopt their method, we can address the different position angle between the inner disk and the line connecting the shadows.  Furthermore, the inclination of the inner disk can also be constrained to close to -45 $^{\circ}$ case in contrast with 45 $^{\circ}$ case. This is because the line connecting two dips(85 $^{\circ}$ and 225 $^{\circ}$) seen in the NIR scattered light image lies on the the south side of the stellar position.  This picture is close to the case ($\theta_1$=-45$^{\circ}$, $\theta_2$=0$^{\circ}$) shown in the second left panel on the bottom row of figure 2 in \citet{2017A&A...604L..10M} in contrast with the case ($\theta_1$=45$^{\circ}$, $\theta_2$=0$^{\circ}$) shown in the second right panel on the bottom row of the same figure. This constrained inclination requires that the northern portion of the inner disk be on the far side while the southern portion is on the near side. In this case, the inner disk rotation is counterclockwise combined with the figure \ref{fig1}(e) and (f).

\subsection{Comparisons with previous observations} \label{sect:cycle2}

The highlights of cycle 3 are detection of the inner disk and twisted butterfly kinematic pattern. The most fundamental difference  between cycle 2 and cycle 3 is that the cycle 3 dataset has a much longer integration time (1.5 hours vs 5 minutes in the cycle 2 dataset), thus reaching a much higher sensitivity level. 
The cycle 2 dataset didn't detect the above-mentioned new structures due to low S/N.

Among disks with symmetric pairs of azimuthal shadows (e.g.,HD 100543, DoAr 44, and HD 142527 \citep{benisty17, 2018MNRAS.477.5104C}), only HD 142527 and  J1604-2130 have been confirmed to have a highly misaligned inner disk via the detection of the twisted butterfly pattern in their gas moment 1 maps. It is interesting to note the difference between the two systems. 

In the gas moment 1 map, both J1604-2130 and HD142527 show a twisted kinematic pattern which in fact not only a misaligned inner disk but also infall can also produce. Although both mechanisms actually work to produce the twist of HD142527 \citep{2015ApJ...811...92C}, \citet{2014ApJ...782...62R} noted a degeneracy between warps and infall for HD142527. In the case of J1604-2130, a misaligned  disk is confirmed to at least partly contribute to producing the twist for the following two reasons. Firstly, there are two symmetrical shadows. Secondly, J1604-2130, as a dipper source, which has the face-on outer disk, requires a mis-aligned inner disk. When discussing the possibility of infall, our new ALMA data detected no significant evidence of inflow penetrating the cavity. Nevertheless,  \citet{2015A&A...584L...4P, ansdell16dipper} reported the complex, aperiodic variability of its optical light curve with dimming by as much as 40\% of the stellar flux in a span of a few days. To Explain this deep variability, infall might contribute to produce both twist and variability to some degree. But still our ALMA new observation do not offer too much new insights why it is variable. It is because our new observation is at one epoch and did not have good enough resolution to resolve any obscuring structures on inner disk which has size probably smaller than 1 AU.

Another point to note is that HD 142527 has a companion (HD 142527 B) at a projected distance of $\sim$0.1 arcsec \citep{biller12, close14, lacour16}, and \citet{price18} proposed that the companion may be responsible for torquing the circumstellar disk and introducing a mutual inclination between that and the circumbinary disk. So far, direct imaging observations have yet to find a companion inside  J1604-2130's cavity \citep{2017A&A...598A..43C}
 We leave the question of the origin of the misaligned inner disk in  J1604-2130 to future investigations.


\acknowledgments

We thank the ALMA telescope staff and operators for making our observations successful. We also would like to thank our referee for carefully reading our manuscript and providing us with suggestions for improving it.
This paper makes use of the following ALMA data: ADS/JAO.ALMA\#2015.1.000888.S.(PI: E. Akiyama). ALMA is a partnership of ESO (representing its member states),
NSF (USA), and NINS (Japan), together with NRC (Canada),
MOST and ASIAA (Taiwan), and KASI (Republic of Korea), in
cooperation with the Republic of Chile. The Joint ALMA
Observatory is operated by ESO, AUI/NRAO, and NAOJ. 
O.P. is funded by the Royal Society via Dorothy Hodgkin Fellowship.  J.M. is supported by the University of Leeds University Research Scholarship. M.T. was supported by MEXT KAKENHI grant number 22000005 is supported by 15H02063 and 18H05442. E. A. is supported by MEXT/JSPS KAKENHI grant No. 17K05399.


\clearpage



\clearpage

\clearpage

\end{document}